\documentclass[conference]{IEEEtran}
\IEEEoverridecommandlockouts
\usepackage{cite}
\usepackage{amsmath,amssymb,amsfonts}
\usepackage{algorithmic}
\usepackage{graphicx}
\usepackage{textcomp}
\usepackage{xcolor}
\def\BibTeX{{\rm B\kern-.05em{\sc i\kern-.025em b}\kern-.08em
    T\kern-.1667em\lower.7ex\hbox{E}\kern-.125emX}}
    
\usepackage{array}
\usepackage{mynotation}
\usepackage{abbreviations}
\usepackage{amsmath}
\usepackage{algorithm}
\usepackage{caption}
\usepackage{subcaption}
\usepackage{pgfplots}
\usepackage{multicol}
\usepackage{makecell}
\usepackage{tikz}

\pgfplotsset{compat=newest}
\pgfplotsset{plot coordinates/math parser=false}

\newlength{\figurewidth}
\newlength{\figureheight}
\newlength{\figurewidthfull}
\newlength{\figureheightfull}
\usepgfplotslibrary{external} 
\usepackage[absolute,showboxes]{textpos}
\TPMargin{2mm}

\usepackage{scalerel}
\usepackage{adjustbox}
\def\decibel{\scalerel{$d$\kern-1.4pt}{\clipbox{2pt 0pt 0pt 0pt}{B}}}

\usepackage[utf8]{inputenc}
\usepackage[T1]{fontenc}

\newcommand{\meas}{z}

\newcommand{\measvec}{\vec{\meas}}

\newcommand{\figref}[1]{Fig.~\ref{#1}}

\newcommand{\copyrightstatement}{
	\begin{textblock}{13.4}(1.25,15)    
		\noindent
		\footnotesize		
		\hspace{1mm}{\copyright 2019 IEEE. Personal use of this material is permitted. Permission from IEEE must be obtained for all other uses, in any current or future media, including reprinting/republishing this material for advertising or promotional purposes, creating new collective works, for resale or redistribution to servers or lists, or reuse of any copyrighted component of this work in other works.}	
	\end{textblock}
}

\begin{document}
	
\copyrightstatement

\title{Extended Target Tracking and Classification Using Neural Networks}

\author{\IEEEauthorblockN{Barkın Tuncer, Murat Kumru, Emre \"{O}zkan}
\IEEEauthorblockA{\textit{Department of Electrical and Electronics Engineering} \\
\textit{Middle East Technical University}\\
Balgat 06800, Ankara, Turkey \\
\{barkin.tuncer, kumru, emreo\} @ metu.edu.tr}
}

\maketitle

\begin{abstract}
Extended target/object tracking (ETT) problem involves tracking objects which potentially generate multiple measurements at a single sensor scan. 
State-of-the-art ETT algorithms can efficiently exploit the available information in these measurements such that they can track the dynamic behaviour of objects and learn their shapes simultaneously. 
Once the shape estimate of an object is formed, it can naturally be utilized by high-level tasks such as classification of the object type. 
In this work, we propose to use a naively deep neural network, which consists of one input, two hidden and one output layers, to classify dynamic objects regarding their shape estimates. 
The proposed method shows superior performance in comparison to a Bayesian classifier for simulation experiments.
\end{abstract}


\section{Introduction}
\label{Int}
Current advances in intelligent systems, automated vehicles and unmanned aerial vehicles brought the necessity of short-range tracking systems.
In contrast to regular long-range counterparts, it is possible to acquire multiple measurements from an object of interest at each instance using short-range sensors. 
Therefore, they enable us to extract valuable information to a greater extent related to the object contour along with the kinematics of the object, e.g., position, velocity and orientation. 
In this regard, extended target/object tracking (ETT) algorithms have provided systematic ways to process these measurements to estimate the kinematic state of the object together with its shape. 
Algorithms presuming simple shape models, such as circle, rectangle, line, are developed in \cite{Circle, Rect, Stick}. 
In this branch of ETT algorithms, the most common approach is to utilize random matrix models, where the target extent is represented by an ellipse \cite{Koch, Feldman, UmutO, lan2012trackinga}. 
In another line of research, approximate non-parametric models are used to describe the target extent. 
These methods can simultaneously track and learn various shapes without assuming predefined extents.
Random hyper-surface models, as suggested in \cite{Baum1, Baum2}, are examples of this class. 
More recently, algorithms relying on a Gaussian Process (GP) representation of the unknown target extent have been suggested in \cite{EoGp1, EoGp2, EoGp3}. 

Classification of objects while tracking has been a long-standing problem in the literature. 
Various algorithms have been proposed to tackle the identification of targets based on their dynamic behavior, motion cues, attributes, fingerprints, etc. \cite{bar11tracking, blackman99, gordon2002efficient}. 
One of the early works on \textit{joint target tracking and classification} (JTC) was presented in \cite{challa2001joint}. 
Their method aims to compute the joint target state-class posterior density and allows for cross-coupled feedback between state and class. 
In \cite{gordon2002efficient}, authors proposed a particle filter based method, which covers the state and feature space, designed for each class. 
In \cite{smets2007kalman}, authors presented an approach to JTC problem based on belief functions. 
In another fold of studies, \cite{li2007optimal} proposed a batch iterative optimization algorithm which minimizes the Bayes risk involving classification and estimation errors. 
A recursive version of this method is introduced in \cite{liu2011recursive}. 
Recently, the authors of \cite{cao2018extended} also tackled the extended object classification problem by relying on a Bayes risk.

In most of the aforementioned methods, there exists a connection between the classification result and the tracking filter, e.g., the result of the classification manipulates the tracking filter. 
On the contrary, the method proposed in this study considers a rather weak coupling between the tracking and classification tasks, since the output of the tracker is used for classification purposes while there is no established feedback mechanism.

\begin{figure*}[t]
	\centering
	\makebox[0pt]{\includegraphics[trim= 0 130 40 0,clip,scale=0.62]{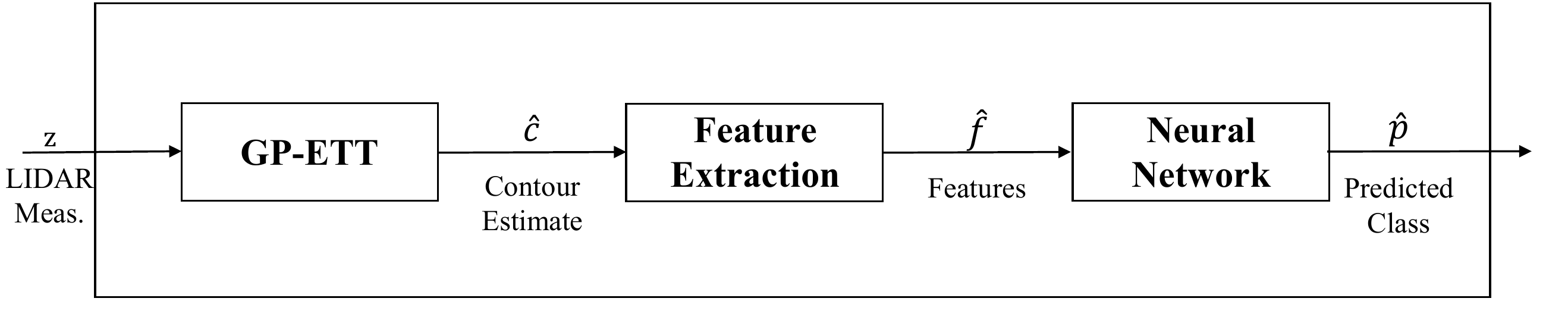}}
	\caption{Block diagram of the proposed online classification algorithm.}
	\label{fig:block1}
\end{figure*}

In this study, we consider the problem of online classification of dynamic objects using point cloud data. 
A block diagram of the proposed method is depicted in Figure \ref{fig:block1}. 
At each instant, point measurements originated from an object are first processed by GP based extended target tracker (GP-ETT) algorithm which produces estimates of the kinematic state and the contour of the object. 
After that, the estimated contour is utilized to extract descriptive features of the object shape. 
These features are then fed to a neural network (NN) model to compute the class probabilities of the object of interest. 
Note that we recently addressed the problem of tracking and classification in \cite{tuncer2018extended}. 
The method in \cite{tuncer2018extended} essentially relies on a similar architecture which exploits the outputs of GP-ETT for classification; however, it carries out the classification by a Bayesian classifier. 
Therefore, the contribution of the current study is twofold: retaining the basic structure of our previous study we hereby demonstrate the modularity of the proposed framework, and secondly, we also improve the resulting classification accuracy by utilizing an NN-based classifier.


As a final note, although GP-ETT is employed as the tracker in this work, it can be simply generalized to other ETT algorithms that are capable of generating extent estimates of the objects. 

The organization of the paper is as follows. 
In Section \ref{sec:ObjectClassification}, we discuss the feature selection process and the details of the NN model. 
Subsequently, we briefly introduce the GP-ETT algorithm in Section \ref{sec:GPETT} to provide an insight into the contour representation that we rely on for classification. 
This is followed by the demonstration of the classification performance via simulation results in Section \ref{simu_results}. 
Finally, we conclude the paper in Section \ref{conclusion}. 


\section{Object Classification} \label{sec:ObjectClassification}
In this work, we aim at achieving object classification by using contour estimates produced by an extended object tracker. 
With this purpose, the estimated contour is to be transformed into some features to acquire a descriptive representation of the object shape. 
Thereafter, an NN-based classifier will perform the classification task regarding these features. 
In this section, the details of the selected features and the classifier architecture are discussed. 

\subsection{Feature Selection for Contour Representation}
In literature, studies on shape classification mostly rely on two different interpretations of the object shape. 
These interpretations are embodied by either region-based descriptors \cite{kim2000region} or contour-based descriptors \cite{latecki2000shape}, \cite{bober2001mpeg}. 
Considering substantial amount of empirical evidence, an object can be described as a combination of a set of regions or may be a single body. 
These regions might include some holes inside. 
Region-based descriptors make use of all information constituting the shape including the holes or several disjoint regions. 
On the other hand, contour-based descriptors consider the characteristic shape features extracted from the contour of an object while ignoring what is inside the contour. 
In this study, we naturally direct our attention to the utilization of the contour-based descriptors since we make use of contour estimates obtained by the tracking algorithm and none of the shapes consist of any holes or disjoint members. 

There are various contour descriptors proposed in the literature. 
The most common ones are the \textit{Fourier Descriptors} (FD), \cite{wallace1980efficient, persoon1977shape}, and  \textit{Curvature Scale Space} (CSS), \cite{mokhtarian1986scale, mokhtarian1992theory}. 

FD consist of the Fourier coefficients of 2-D shapes which can be represented in terms of different shape signatures, such as the contour coordinates expressed as complex numbers or the radial distance between the contour and the object center, \cite{wallace1980efficient}. 
FD are robust to rotation and affine transformation, while being efficient in terms of computation. 
CSS, on the other hand, considers the inflection points as descriptors which represent the location of change in the direction of the curvature. 
The contour is convolved with multiple Gaussian kernels with different standard deviations, also called as width, to smoothen the contour. 
The inflection points are calculated for different Gaussian kernels and used as descriptors.
One of the main advantages of CSS is that it is noise invariant since the characteristic inflection points remain available after filtering with large width kernels, while other points resolve.

In addition to these prominent descriptors, there are also numerous geometric features with various levels of complexity, \cite{yang2017evaluating}, \cite{yang2008survey}, \cite{tao2007preliminary}. 
These features are basically employed to map object contours into some descriptive representations. 
Selecting a proper set of features among numerous alternatives is of paramount importance for classification performance. 
In this regard, we restrict our scope to the simple geometric features to be able to form a basis for computationally efficient and fast operation. 
The following set of six features are chosen to be respected by the classifier: \textit{elongation}\cite{yang2008survey}, \textit{rectangularity}\cite{yang2017evaluating}, \textit{circularity}\cite{yang2017evaluating}, \textit{solidity}\cite{yang2017evaluating}, \textit{compactness}\cite{tao2007preliminary}, and \textit{area}.
Note that this particular group of features is not hand-crafted to optimize the resulting classification performance. 
Instead, it is empirically observed to be sufficient for the proof of concept. 
However, for a specific application, it can be selected to account for any prior knowledge or can be learned in an automated fashion by employing various tools, such as convolutional neural networks, autoencoders. 

\begin{figure}[h!] 
	\centering
	\includegraphics[width = 0.6\columnwidth] {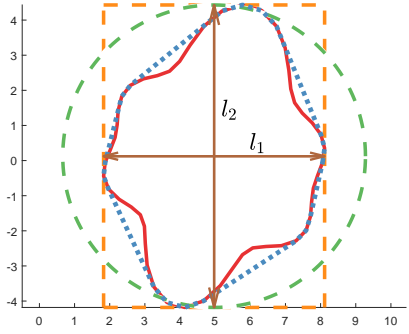}
	\caption{Visualization of the parameters used in the feature extraction process. 
	The solid red curve represents the contour of the object. 
	The minimum bounding circle and rectangle are plotted in green and yellow, respectively. 
	Major and minor axes of the minimum bounding rectangle are denoted by $l_1$ and $l_2$. 
	The blue dashed line indicates the convex hull of the object.}
	\label{fig:features}
\end{figure}

Elongation is uniquely defined as the ratio of the major and minor axes of the minimum bounding rectangle (which are denoted as $l_1$ and $l_2$ in \figref{fig:features}). 
Rectangularity is the ratio of the object area to the area of the minimum size rectangle that encloses the contour. 
In \figref{fig:features}, rectangularity can be computed as the object area to the area encapsulated by the orange contour. 
On the other hand, circularity is related to the ratio of the object area to the area of the minimum confining circle, and it is basically used to measure the similarity of the object of interest to a circle. 
The minimum bounding circle is illustrated by the green line in \figref{fig:features}, and the circularity is calculated as the ratio of the object area to the area enclosed by the green contour. 
Solidity is an indicator of the shape being convex or concave. 
It is defined as the ratio of the object area to the area of the convex hull. 
The convex hull of the shape is represented by the blue line in \figref{fig:features}. 
Lastly, compactness measures the contour complexity versus the enclosed area. 
The value of compactness increases with increasing shape complexity. 
The expression to calculate the compactness is given as \cite{tao2007preliminary}
\begin{equation}
\label{eq:compactness}
    C = 1 - \frac{4{\pi}a}{p^2},
\end{equation}
where $a$ and $p$ are the area and the perimeter of the shape, respectively. 

The computed features of the object contour are then passed to the NN-based classifier. 
The classifier architecture is revealed in the following subsection.

\subsection{Classifier Architecture}
\begin{figure}[h!] 
	\centering
	\includegraphics[trim= 0 0 0 0,clip, width = 0.75\columnwidth] {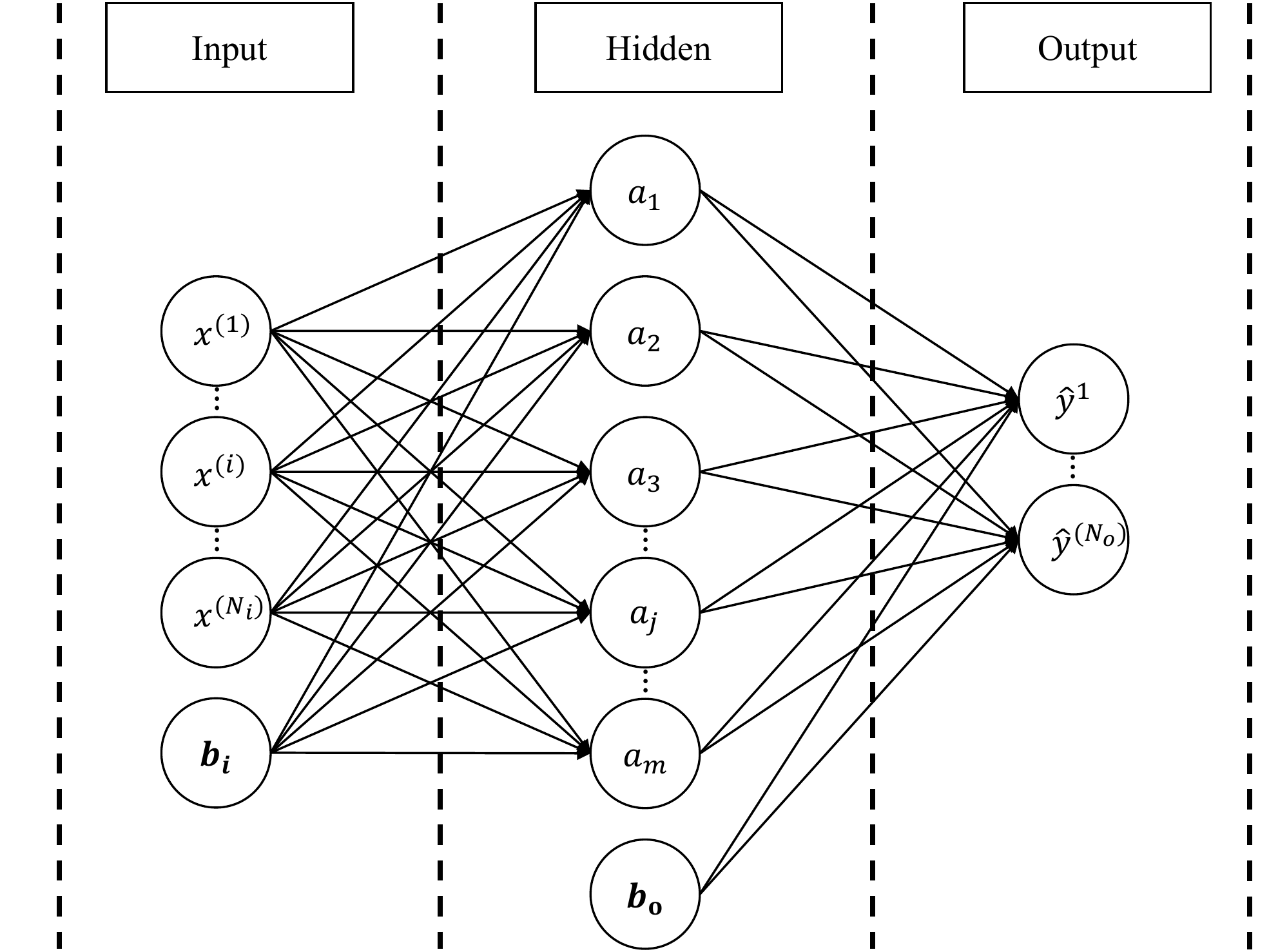}
	\caption{A simple neural network including one input layer, one hidden layer and one output layer.}
	\label{fig:neural_network}
\end{figure}

In this study, we realize object classification by a naively deep feedforward NN. 
A typical NN model is basically a collection of special processing units, called neurons, which are grouped into different layers, such as input, hidden and output layers. 
Specifically, in a feedforward NN, there exist weighted connections between these layers while there is no connection between neurons within the same layer.
A simple feedforward NN consisting of one input, one hidden and one output layer is depicted in \figref{fig:neural_network}. 

In vector form, the outputs of the neurons in the hidden layer can be computed as follows. 
\begin{equation} \label{eq:NNHiddenOut}
\mathbf{a}  = \mathbf{g}\left(\mathbf{W}_i^\Transp \mathbf{x}+ \mathbf{b}_i\right)
\end{equation}
$\textbf{a} \triangleq [a_1 \dots a_m]^\Transp$ represents the output vector of the hidden layer; $\textbf{x} \triangleq [x^{(1)} \dots x^{(N_i)}]^\Transp$ indicates the input vector of the network; $\textbf{W}_i$ is the input weight matrix; $\textbf{b}_i \triangleq [b_1 \dots b_m]^\Transp$ 
denotes the bias vector; $\mathbf{g}(\cdot)$ is the activation function. 
Note that if the NN model has multiple hidden layers, the output of $k^{th}$ hidden layer can easily be calculated simiilarly via replacing $\textbf{x}$ in \eqref{eq:NNHiddenOut} by the output of the $(k-1)^{th}$ layer. 

Besides, the output of the network is obtained by
\begin{equation} 
\hat{\mathbf{y}}  = \mathbf{W}_o^\Transp \mathbf{a} + \mathbf{b}_o
\end{equation}
where $\hat{\mathbf{y}} \triangleq [\hat{y}_1 \dots \hat{y}_{N_o}]^\Transp$ is the output vector; $\mathbf{W}_o$ is the output weight matrix and $\mathbf{b}_o$ is the bias vector. 

The standard loss function used in the training procedure to optimize the parameters of an NN is given in \eqref{eq:total_loss1}. 
\begin{align}
\label{eq:total_loss1}
	J(\theta)  = \frac{1}{N}\sum\limits_{i=1}^{N} \left(  y^{(i)} - f_\theta(x^{(i)}) \right)^2     
\end{align}
There are $N$ training points and $y^{(i)}$ denotes the ground truth for the $i^{th}$ input $x^{(i)}$. 
$f_\theta(x^{(i)})$ is the output of the NN for the corresponding input, and it is parametrized by $\theta$ which comprises of the weights and the biases of the model. 


In this work, we construct a naively deep NN consisting of 2 hidden layers with respectively 16 and 8 neurons. 
This specific architecture was empirically observed to be sufficient regarding the classification performance obtained for training and validation data sets. 
The activation functions are selected as \textit{hyperbolic tangent sigmoid} function in the hidden layers and \textit{softmax} function in the output layer. 

\section{GP-ETT}
\label{sec:GPETT}
First introduced in \cite{EoGp1}, GP-ETT is an effective way of tracking dynamic objects with unknown shapes. 
It is able to jointly estimate the kinematics and the extent of the object by using point measurements. 
The suggested NN-based object classification scheme basically processes the contour estimates produced by this algorithm. 

\begin{figure}
	\centering
	\begin{subfigure}{0.42\columnwidth}
		\centering
		\includegraphics[trim= 205 275 190 290,clip, scale=0.5] {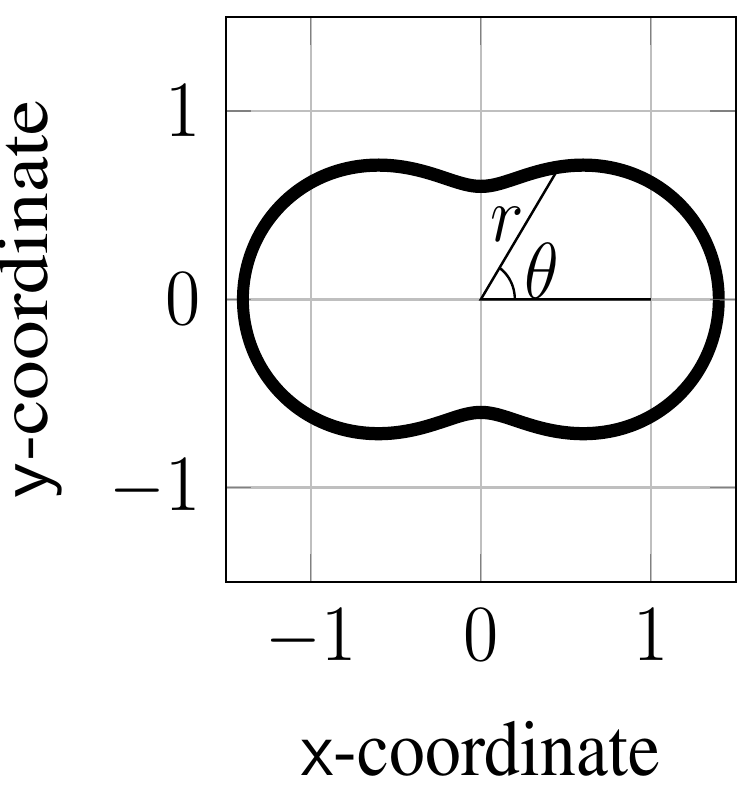}
		\caption{Star-convex target shape}
		\label{fig:intro2}
	\end{subfigure}
	\hspace{3mm}
	\begin{subfigure}{0.42\columnwidth}
		\centering
		\includegraphics[trim= 200 275 190 290,clip, scale=0.5] {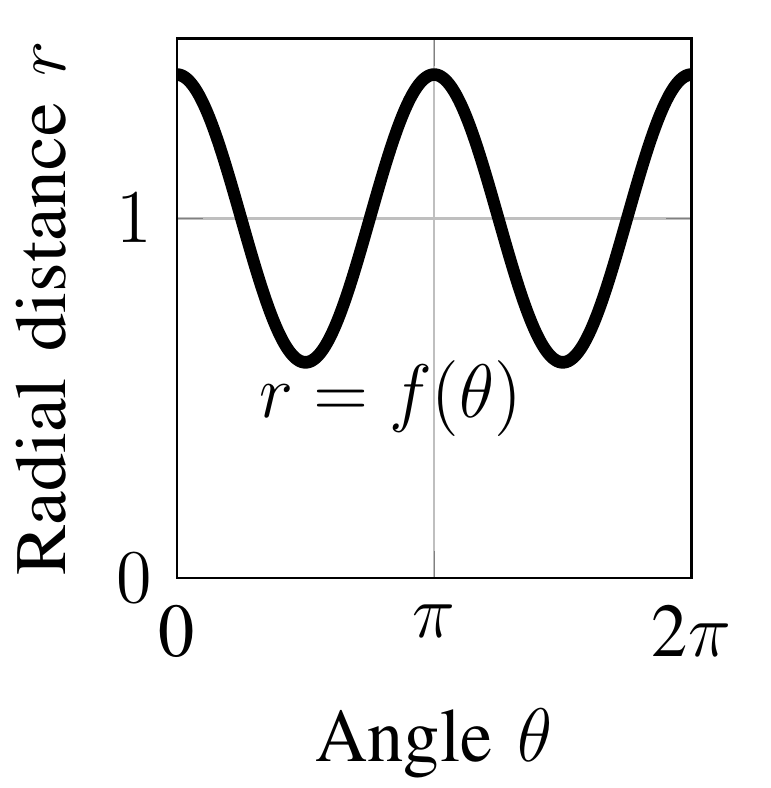}
		\caption{Radial function}
		\label{fig:intro1}
	\end{subfigure}
	\caption{An example star-convex contour described by radial function $r=f(\theta)$, \cite{EoGp1}.}
	\label{fig:intro}
\end{figure}

In the formulation of GP-ETT, the contour of an object is represented by a radial function $r=f(\theta)$ which is to be modeled by a GP. 
The output of the radial function is the distance between the center and the contour at the specified polar angle. 
A typical shape described by radial function is shown in \figref{fig:intro}. 
Notice that the formulation implicitly assumes that the shape of the object is \emph{star-convex}\footnote{A set $\mathcal{S}(x)$ is called star-convex if each line segment from the center to any point is fully contained in $\mathcal{S}(x)$, where $x$ denotes the position of a point.}.

A noisy observation of the object contour represented by the radial function can be described as 
\begin{equation} \label{eq:meas_model1}
\measvec_{k,l} = \pos_k +  \p(\theta_{k,l})f(\theta_{k,l}) + \e_{k,l},
\end{equation}
where $\pos_k$ is the center position of the target at time instant $k$, 
$\{\measvec_{k,l}\}_{l=1}^{n_k}$ indicate the measurements collected at time $k$, 
$\{\theta_{k,l}\}_{l=1}^{n_k}$ are the polar angles of the source points on the contour that originate the corresponding measurements, $\e_{k,l} \sim \Nd{0}{\R}$ denote i.i.d. Gaussian noise with zero mean and covariance \textit{R}, 
and $\p(\theta_{k,l})$ is an orientation vector defined as
${\p(\theta_{k,l}) \triangleq \begin{bmatrix} \cos(\theta_{k,l}) \\ \sin(\theta_{k,l}) \end{bmatrix}^\Transp}$.

\begin{figure}[h!] 
	\centering
%
%
\begin{tikzpicture}

\begin{axis}[%
xlabel= Angle Difference $(\theta - \theta')$,
ylabel= Covariance,
width=2.674in,
height=1.8in,
at={(0.448in,0.285in)},
scale only axis,
xmin=-0.2,
xmax=6.48318530717959,
xtick={0,3.14159265358979,6.28318530717959},
xticklabels={{$0$},{$\pi$},{$2\pi$}},
tick align=outside,
ymin=0,
ymax=6,
ytick={1,5},
yticklabels={{$\sigma_r^2$},{$\sigma_f^2 + \sigma_r^2$}},
axis background/.style={fill=white},
axis x line*=bottom,
axis y line*=left,
xmajorgrids,
ymajorgrids,
grid style={dotted}
]
\addplot [color=black!60!teal, line width=2.5pt, forget plot]
  table[row sep=crcr]{%
0	5\\
0.0315737955134653	4.99676933225539\\
0.0631475910269305	4.98709618772129\\
0.0947213865403958	4.97103693542018\\
0.126295182053861	4.94868483696938\\
0.157868977567326	4.92016902862719\\
0.189442773080792	4.88565312021576\\
0.221016568594257	4.84533343430162\\
0.252590364107722	4.79943691475107\\
0.284164159621187	4.74821873884484\\
0.315737955134653	4.69195967142992\\
0.347311750648118	4.63096320302133\\
0.378885546161583	4.565552516286\\
0.410459341675048	4.49606732691085\\
0.442033137188514	4.42286064547258\\
0.473606932701979	4.34629550660621\\
0.505180728215444	4.26674171055611\\
0.536754523728909	4.18457262015191\\
0.568328319242375	4.10016205346626\\
0.59990211475584	4.01388130897996\\
0.631475910269305	3.92609635611342\\
0.66304970578277	3.83716521960022\\
0.694623501296236	3.7474355815001\\
0.726197296809701	3.6572426197976\\
0.757771092323166	3.56690709762695\\
0.789344887836631	3.47673371231561\\
0.820918683350097	3.38700970875062\\
0.852492478863562	3.29800375713352\\
0.884066274377027	3.20996509107959\\
0.915640069890493	3.12312289829723\\
0.947213865403958	3.03768595280218\\
0.978787660917423	2.95384247481191\\
1.01036145643089	2.87176020214553\\
1.04193525194435	2.79158665513013\\
1.07350904745782	2.71344957567637\\
1.10508284297128	2.63745752031718\\
1.13665663848475	2.56370058657418\\
1.16823043399821	2.49225125199275\\
1.19980422951168	2.42316530552573\\
1.23137802502514	2.3564828516038\\
1.26295182053861	2.29222936815769\\
1.29452561605208	2.23041680100759\\
1.32609941156554	2.17104467835803\\
1.35767320707901	2.11410123058715\\
1.38924700259247	2.05956450205422\\
1.42082079810594	2.00740344322731\\
1.4523945936194	1.9575789730197\\
1.48396838913287	1.91004500278618\\
1.51554218464633	1.86474941494267\\
1.5471159801598	1.82163499061226\\
1.57868977567326	1.78064028204969\\
1.61026357118673	1.74170042684256\\
1.64183736670019	1.70474790201979\\
1.67341116221366	1.66971321721333\\
1.70498495772712	1.63652554691277\\
1.73655875324059	1.60511330262739\\
1.76813254875405	1.57540464642773\\
1.79970634426752	1.54732794788462\\
1.83128013978099	1.52081218686444\\
1.86285393529445	1.49578730498207\\
1.89442773080792	1.47218450876716\\
1.92600152632138	1.44993652777287\\
1.95757532183485	1.4289778309585\\
1.98914911734831	1.40924480471832\\
2.02072291286178	1.39067589591553\\
2.05229670837524	1.37321172322317\\
2.08387050388871	1.35679515997887\\
2.11544429940217	1.34137139163645\\
2.14701809491564	1.32688795075048\\
2.1785918904291	1.31329473226583\\
2.21016568594257	1.30054399170897\\
2.24173948145603	1.2885903286954\\
2.2733132769695	1.27739065798209\\
2.30488707248296	1.26690417010898\\
2.33646086799643	1.25709228349172\\
2.36803466350989	1.24791858965134\\
2.39960845902336	1.23934879309742\\
2.43118225453682	1.23135064722054\\
2.46275605005029	1.22389388739871\\
2.49432984556376	1.21695016238106\\
2.52590364107722	1.21049296488192\\
2.55747743659069	1.20449756219803\\
2.58905123210415	1.19894092755249\\
2.62062502761762	1.19380167276982\\
2.65219882313108	1.18905998279736\\
2.68377261864455	1.18469755250904\\
2.71534641415801	1.1806975261564\\
2.74692020967148	1.17704443977052\\
2.77849400518494	1.17372416676366\\
2.81006780069841	1.1707238669331\\
2.84164159621187	1.16803193902918\\
2.87321539172534	1.16563797701571\\
2.9047891872388	1.16353273012204\\
2.93636298275227	1.16170806676281\\
2.96793677826573	1.16015694238167\\
2.9995105737792	1.15887337126005\\
3.03108436929266	1.15785240231998\\
3.06265816480613	1.15709009894049\\
3.0942319603196	1.15658352280047\\
3.12580575583306	1.15633072175586\\
3.15737955134653	1.15633072175586\\
3.18895334685999	1.15658352280047\\
3.22052714237346	1.15709009894049\\
3.25210093788692	1.15785240231998\\
3.28367473340039	1.15887337126005\\
3.31524852891385	1.16015694238167\\
3.34682232442732	1.16170806676281\\
3.37839611994078	1.16353273012204\\
3.40996991545425	1.16563797701571\\
3.44154371096771	1.16803193902918\\
3.47311750648118	1.1707238669331\\
3.50469130199464	1.17372416676366\\
3.53626509750811	1.17704443977052\\
3.56783889302157	1.1806975261564\\
3.59941268853504	1.18469755250904\\
3.6309864840485	1.18905998279736\\
3.66256027956197	1.19380167276982\\
3.69413407507544	1.19894092755249\\
3.7257078705889	1.20449756219803\\
3.75728166610237	1.21049296488192\\
3.78885546161583	1.21695016238106\\
3.8204292571293	1.22389388739871\\
3.85200305264276	1.23135064722054\\
3.88357684815623	1.23934879309742\\
3.91515064366969	1.24791858965134\\
3.94672443918316	1.25709228349172\\
3.97829823469662	1.26690417010898\\
4.00987203021009	1.27739065798209\\
4.04144582572355	1.2885903286954\\
4.07301962123702	1.30054399170897\\
4.10459341675048	1.31329473226583\\
4.13616721226395	1.32688795075048\\
4.16774100777741	1.34137139163645\\
4.19931480329088	1.35679515997887\\
4.23088859880434	1.37321172322317\\
4.26246239431781	1.39067589591553\\
4.29403618983128	1.40924480471832\\
4.32560998534474	1.4289778309585\\
4.35718378085821	1.44993652777287\\
4.38875757637167	1.47218450876716\\
4.42033137188514	1.49578730498207\\
4.4519051673986	1.52081218686444\\
4.48347896291207	1.54732794788462\\
4.51505275842553	1.57540464642773\\
4.546626553939	1.60511330262739\\
4.57820034945246	1.63652554691277\\
4.60977414496593	1.66971321721333\\
4.64134794047939	1.70474790201979\\
4.67292173599286	1.74170042684256\\
4.70449553150632	1.78064028204969\\
4.73606932701979	1.82163499061226\\
4.76764312253325	1.86474941494267\\
4.79921691804672	1.91004500278618\\
4.83079071356018	1.9575789730197\\
4.86236450907365	2.00740344322731\\
4.89393830458711	2.05956450205422\\
4.92551210010058	2.11410123058715\\
4.95708589561405	2.17104467835803\\
4.98865969112751	2.23041680100759\\
5.02023348664098	2.29222936815769\\
5.05180728215444	2.3564828516038\\
5.08338107766791	2.42316530552573\\
5.11495487318137	2.49225125199275\\
5.14652866869484	2.56370058657418\\
5.1781024642083	2.63745752031718\\
5.20967625972177	2.71344957567637\\
5.24125005523523	2.79158665513013\\
5.2728238507487	2.87176020214553\\
5.30439764626216	2.95384247481191\\
5.33597144177563	3.03768595280217\\
5.36754523728909	3.12312289829723\\
5.39911903280256	3.20996509107959\\
5.43069282831602	3.29800375713352\\
5.46226662382949	3.38700970875062\\
5.49384041934295	3.47673371231561\\
5.52541421485642	3.56690709762695\\
5.55698801036988	3.6572426197976\\
5.58856180588335	3.7474355815001\\
5.62013560139682	3.83716521960022\\
5.65170939691028	3.92609635611342\\
5.68328319242375	4.01388130897996\\
5.71485698793721	4.10016205346626\\
5.74643078345068	4.18457262015191\\
5.77800457896414	4.26674171055611\\
5.80957837447761	4.34629550660621\\
5.84115216999107	4.42286064547257\\
5.87272596550454	4.49606732691085\\
5.904299761018	4.565552516286\\
5.93587355653147	4.63096320302133\\
5.96744735204493	4.69195967142992\\
5.9990211475584	4.74821873884484\\
6.03059494307186	4.79943691475107\\
6.06216873858533	4.84533343430162\\
6.09374253409879	4.88565312021576\\
6.12531632961226	4.92016902862719\\
6.15689012512572	4.94868483696938\\
6.18846392063919	4.97103693542018\\
6.22003771615265	4.98709618772129\\
6.25161151166612	4.99676933225539\\
6.28318530717959	5\\
};
\end{axis}
\end{tikzpicture}%
	\caption{The covariance function of the GP as defined in \eqref{eq:SEcov1}.}
	\label{fig:covariance_function}
\end{figure}
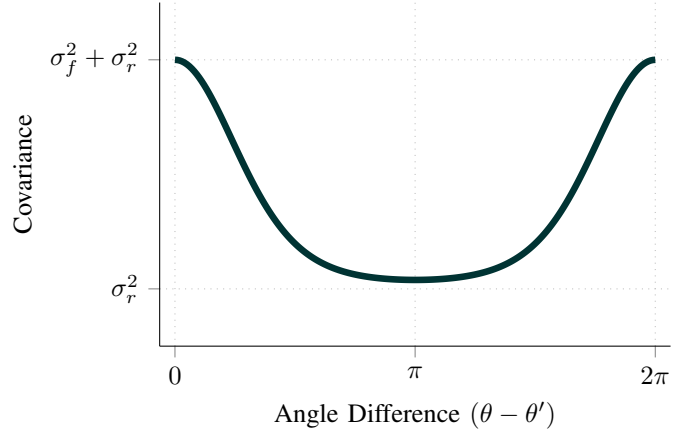

The core idea of GP-ETT is to facilitate modeling of the unknown radial function $f(\cdot)$ in \eqref{eq:meas_model1} by a GP, i.e., ${f(\theta) \sim \GP\big(\mu(\theta),k(\theta,\theta')\big)}$ where $\mu(\cdot)$ is the mean function and $k(\cdot, \cdot)$ is the covariance function.
By doing so, one can establish a probabilistic representation of the extent whose inherent spatial characteristics are conveniently encoded by the covariance function. 
In this work, the GP is specified with the zero mean function and the following covariance function, which is obtained by modifying the squared exponential kernel.  
\begin{equation} \label{eq:SEcov1}
k(\theta,\theta') = \sigma_f^2 e^{-\frac{2\text{sin}^2 \left( \frac{\lvert \theta-\theta' \rvert}{2} \right) }{l^2}} + \sigma_r^2
\end{equation}
$\sigma_f$ is the prior variance, $l$ is the scaled length and $\sigma_r$ essentially accounts for the uncertainty in the mean. 
The covariance function is illustrated in Fig. \ref{fig:covariance_function}. 
Being a stationary covariance function which depends on the relative position of the inputs, the function is plotted against the difference between the input angles, i.e., $\theta - \theta'$. 
The figure reveals that the covariance function ensures perfect correlation between $f(\theta)$ and $f(\theta+2\pi)$, i.e., ${\rho \left( f(\theta), f(\theta') \right) =  1}$, since they basically correspond to the same point on the contour. 
Additionally, the correlation between $f(\theta)$ and $f(\theta')$ decreases as the angle between them increases. 
This characteristic essentially accounts for the fact that radii at closer angles naturally tend to be more interrelated than the farther sections of the contour. 

The standard GP regression requires batch processing of the measurements hence it does not apply to the online tracking and classification application due to its computational complexity and sequentially available measurements. 
To overcome these issues, an approximation of the standard regression was suggested to compute the posterior recursively in \cite{EoGp1}. 
In particular, this method summarizes the original GP model at a finite number of basis inputs ${\mathbf{u} = \begin{bmatrix} u_{1} & \dots & u_{N} \end{bmatrix}^\Transp}$, thus the posterior distribution can simply be computed by a Kalman filter regarding the following state space model. 
\begin{subequations} 
	\label{eq:xf}
	\begin{align}
	\mathbf{c}_{k+1}  &= \F \mathbf{c_{k}} + \mathbf{w}_{k}, \quad \mathbf{w}_{k} \sim \mathcal{N}(0, {Q}), \\
	\meas_{k} &= H(u_k)\mathbf{c}_{k} + \mathbf{e}, \quad \mathbf{e} \sim \mathcal{N}(0, \R(u_k)), \\
	\mathbf{c}_0  &\sim \mathcal{N}(0,P_0),
	\end{align}
\end{subequations}
where $\mathbf{c}_k = \begin{bmatrix} f(u_{1}) & \dots & f(u_{N}) \end{bmatrix}^\Transp$ is the radial function values at time $k$, and $z_k$ indicates a single point measurement. 
Further details of the model can be found in \cite{EoGp1}.

To be able jointly estimate the target dynamics and the contour, a unified state space model is constructed which relies on the state vector, $\textbf{x}_k \triangleq [\bar{\textbf{x}}_k \ \  \textbf{c}_k]$, where $\bar{\textbf{x}}_k$ denotes the object dynamics including position, velocity and orientation. 
GP-ETT algorithm is simply implemented by an extended Kalman filter (EKF) considering this state vector. 
The EKF infers the joint posterior of the state, $p(\textbf{x}_k | \textbf{z}_{1:k})$, and the updated mean of the contour  $\hat{c}$ is delivered to the classification scheme as the contour estimate. 

As GP-ETT recursively updates the posterior of the state, it thereby refines the contour estimate due to the accumulation of information in time. 
In contrast, instant point measurements can only delineate the shape partially, and they get sparse due to occlusions or increased distance between the object and the sensor. 
Consequently, GP-ETT provides a more reliable basis to perform classification compared to instant measurements. 


\section{Simulations and Results} \label{simu_results}
\begin{figure}[h!] 
	\centering
	\includegraphics[trim= 0 0 0 0,clip, width = \columnwidth] {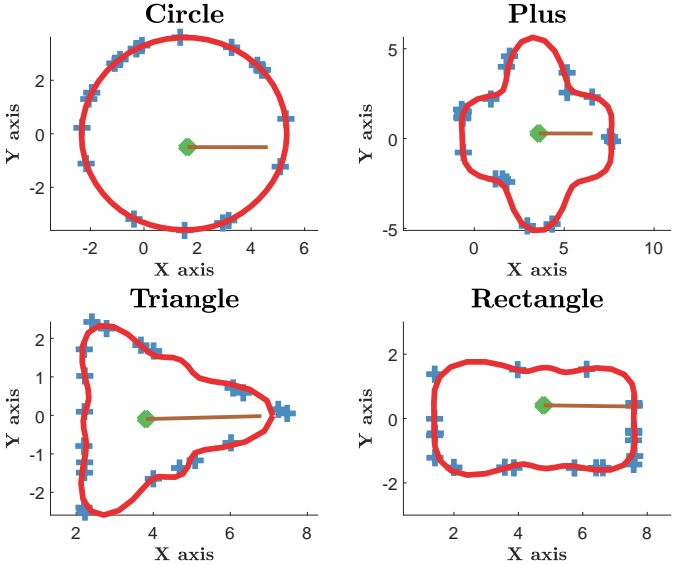}
	\caption{Typical contour estimates produced by GP-ETT. (Contour estimates are plotted in blue while red crosses represent point measurements.)}
	\label{fig:shapes}
\end{figure}

In this section, the performance of the proposed algorithm is demonstrated through some simulation experiments. 
Throughout simulations, objects from the following shape classes are taken into consideration: circle, triangle, rectangle and plus. 
To form a labeled data set, first we simulate dynamic scenarios of random sized objects from each class. 
The initial orientation of each object is randomly selected from a uniform distribution between 0 and $2\pi$. 
At each instant of the simulations, point measurements are originated from random points on the object contour with an additive Gaussian noise whose standard deviation is set to 0.02. 
Subsequently, these point measurements are processed by the GP-ETT to obtain corresponding contour estimates. 
Contour estimates are maintained at fifty basis angles which are evenly spaced between $0$-$2\pi$. 
Some typical examples of the contour estimates from each class are exhibited in \figref{fig:shapes}. 
The resulting data set consists of 10000 contour estimates per each shape class, and it is divided into training and test sets by the ratios of 80\% and 20\%, respectively.
The training set is further split into two by the same ratios, 80\% and 20\%, to attain training and validation sets. 

To be able to assess the performance of the proposed algorithm in a comparative manner, we consider another classification method from our previous work, \cite{tuncer2018extended}. 
This baseline algorithm is based on a Bayesian classification scheme and is denoted as `BC'. 
Different from the proposed method, BC makes use of both the contour estimate and the associated covariance matrix, which essentially carries the local uncertainty information along the contour estimate. 
In particular, BC applies Unscented Transform (UT) to the outputs of GP-ETT to adopt a probabilistic representation in the feature space. 
Subsequently, the classification is achieved by the following Bayesian classifier which regards this probabilistic description of the feature.
\begin{align}
\label{eqn:bayes}
Pr(r=i|f) =\frac{ p(f|r=i)p(r=i) }{\sum_{j=1}^M  p(f|r=j)p(r=j)} 
\end{align}
where $f$ is the feature vector; ${r \in \{1, \ldots, M\}}$ is the class index, and $p(f|r=j)$ is the likelihood of observing the feature $f$ from an object in class-$j$. 
The posterior probability, ${Pr(r=i|f)}$, can simply be computed for each class once the sufficient statistics of the class distributions are available. 
To this end, the feature distribution of each class is approximated to be Gaussian, i.e., ${p(f|r=i) \approx  \mathcal{N}(f; \mu_i,\Sigma_i)}$, and the mean, $\mu_i$, and the covariance matrix, $\Sigma_i$, are determined by means of a supervised learning scheme using the labeled training data. 
For further details, see \cite{tuncer2018extended}.  

Additionally, we also implemented a variant of the proposed method which employs an NN having the same architecture; however, this time the NN is trained and tested directly on the contour estimates rather than the features extracted from these estimates. 
The original method is denoted as `NN-feature' and `NN-contour' stands for the variant.

\begin{figure}[h!] 
	\centering
	\includegraphics[width = 0.9\columnwidth] {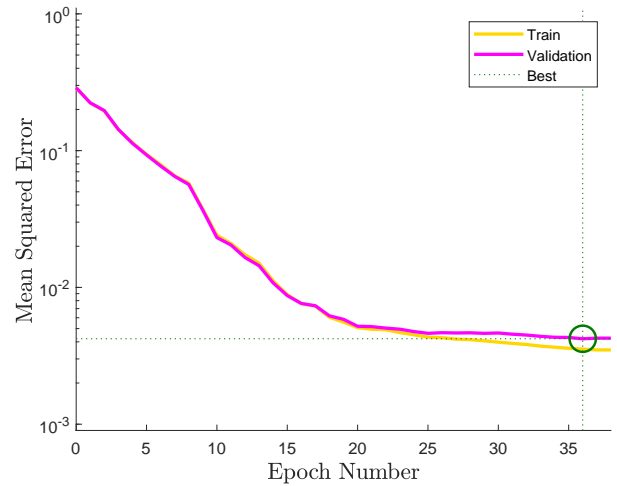}
	\caption{The performance curves of the proposed model (NN-feature). The network parameters are set to be the values satisfying the best performance in the validation set.}
	\label{fig:loss}
\end{figure}

All of the algorithms were implemented in MATLAB 2018a; in particular, we used Deep Learning Toolbox for NN-feature and NN-contour. 
Since the NN model is a rather shallow one, we optimized the network parameters by Levenberg-Marquardt method, which provides a fast but computationally expensive optimization scheme combining the gradient descent algorithm with Gauss-Newton method. 
The training procedure is stopped when the validation accuracy does not increase for the following 20 consecutive epochs. 
The loss function values of NN-feature at each epoch are demonstrated for both training and validation sets in Fig. \ref{fig:loss}. 
The training and test of the models were performed on a computer with an Intel Core i7-6700HQ CPU without using parallel programming. 

The results obtained by the methods are presented in Table \ref{table:table_1}. 
NN-feature is observed to outperform the other algorithms in the accuracy rate. 
Specifically, the difference in the accuracy rates of NN-feature and BC is mainly due to the test scenarios including irregular contour estimates generated due to insufficient sampling of the objects. 
Some typical classification outputs produced at these instants are depicted in Table \ref{tbl:results}. 
NN-feature can handle these challenging scenarios more robustly. 
Even though NN-contour has a higher accuracy rate compared to BC, the execution time per object is significantly higher than the others. 
Particularly, NN-contour operates more slowly compared to NN-feature because the former processes fifty contour points while the latter one considers only six features. 
This characteristic might render NN-contour inconvenient for object classification in real time. 


\begin{table}[b!]
	\centering
	\caption{Classification results on the synthetic data set in terms of accuracy rate and the execution times for each shape.}
	\renewcommand{\arraystretch}{1.5}
	\begin{tabular}{||c c c||} 
		\hline		
		\small{Classifier} & \makecell{ \small{Accuracy} \\ \small{Rate} } &  \makecell{ \small{Execution Time} \\ \small{(ms/object)} } \\		
		\hline\hline		
		\small{BC}& 0.94 & \textbf{0.57} \\		
		\hline		
		\small{NN-feature}& \textbf{0.99} & 1.80 \\		
		\hline	
		\small{NN-contour}& 0.97 & 2.55 \\		
		\hline
	\end{tabular}
	\label{table:table_1}
\end{table}

\begin{table*}[t]
  \caption{Some of the results that NN shows superior performance against Bayesian approach.}\label{tbl:results}
  \centering
  \begin{tabular}{ || p{3.5cm} | p{3.5cm} | p{3.5cm} | p{3.5cm} ||}
    \hline\hline
    \begin{minipage}{.25\textwidth}
      \flushleft
      \hspace{0.5cm}\includegraphics[trim= 0 0 0 -10,clip,scale=0.5]{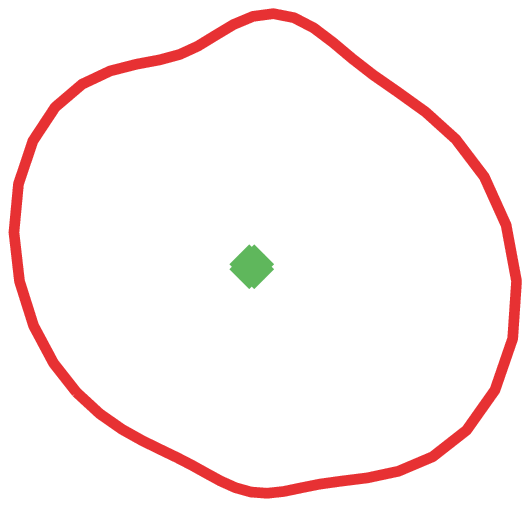}
      \\ 
      \flushleft
      \hspace{0.5cm}Ground Truth: Circle \\ 
      \hspace{0.5cm}Bayesian: Plus \\ 
      \hspace{0.5cm}NN: \textbf{Circle} \\ 
    \end{minipage}
      
    &
    \begin{minipage}{.25\textwidth}
    \flushleft
      \hspace{0.5cm}\includegraphics[trim= 0 0 0 -10,clip,scale=0.5]{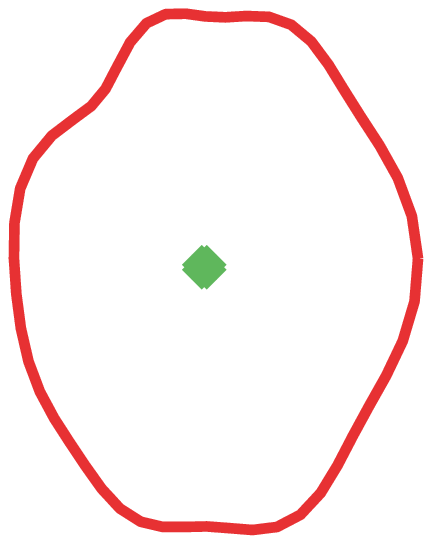}
      \\ 
      \flushleft
      \hspace{0.5cm}Ground Truth: Plus \\ 
      \hspace{0.5cm}Bayesian: Circle \\ 
      \hspace{0.5cm}NN: \textbf{Plus} \\ 
    \end{minipage}
    
    & 
    \begin{minipage}{.25\textwidth}
    \flushleft
      \hspace{0.5cm}\includegraphics[trim= 0 0 0 -10,clip, scale=0.5]{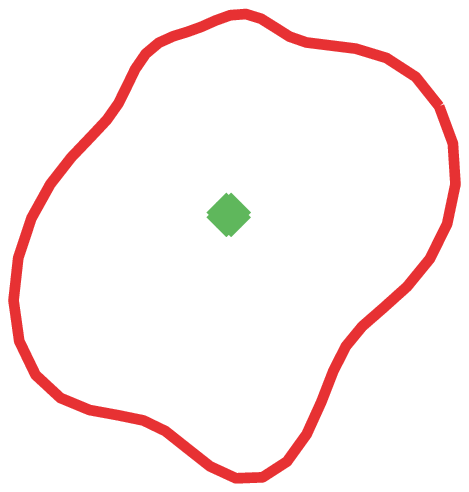}
      \\ 
      \flushleft
      \hspace{0.5cm}Ground Truth: Plus \\ 
      \hspace{0.5cm}Bayesian: Rectangle \\ 
      \hspace{0.5cm}NN: \textbf{Plus} \\ 
    \end{minipage}
    
    & 
    \begin{minipage}{.25\textwidth}
    \flushleft
      \hspace{0.5cm}\includegraphics[trim= 0 0 0 -10,clip,scale=0.5]{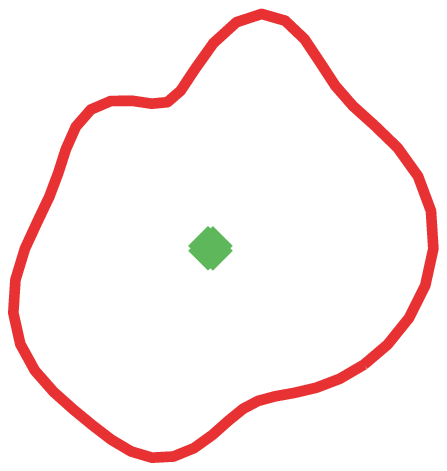}
     \\ 
     \flushleft
     \vspace{0.2cm}
     \hspace{0.5cm}Ground Truth: Plus \\ 
      \hspace{0.5cm}Bayesian: Rectangle \\ 
      \hspace{0.5cm}NN: \textbf{Plus} \\ 
    \end{minipage}
    \\ \hline
    \begin{minipage}{.25\textwidth}
    \flushleft
      \hspace{0.5cm}\includegraphics[trim= 0 0 0 -10,clip,scale=0.5]{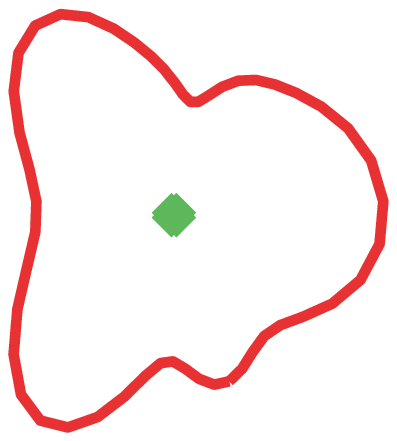}
      \\ 
      \flushleft
      \hspace{0.5cm}Ground Truth: Triangle \\ 
      \hspace{0.5cm}Bayesian: Rectangle \\ 
      \hspace{0.5cm}NN: \textbf{Triangle} \\ 
    \end{minipage}
    &
    \begin{minipage}{.25\textwidth}
    \flushleft
      \hspace{0.5cm}\includegraphics[trim= 0 0 0 -10,clip,scale=0.5]{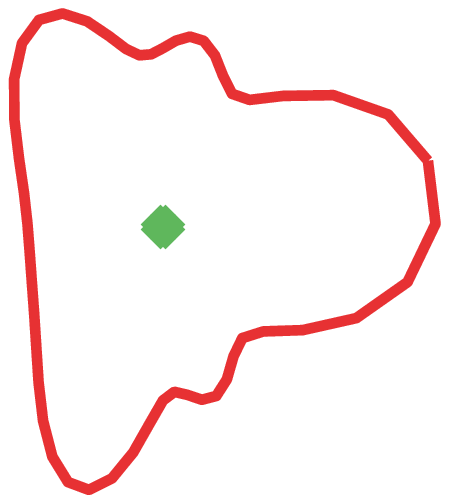}
      \\ 
      \flushleft
     \hspace{0.5cm}Ground Truth: Triangle \\ 
      \hspace{0.5cm}Bayesian: Plus \\ 
      \hspace{0.5cm}NN: \textbf{Triangle} \\ 
    \end{minipage}
    & 
    \begin{minipage}{.25\textwidth}
    \flushleft
      \hspace{0.5cm}\includegraphics[trim= 0 0 0 -10,clip,scale=0.5]{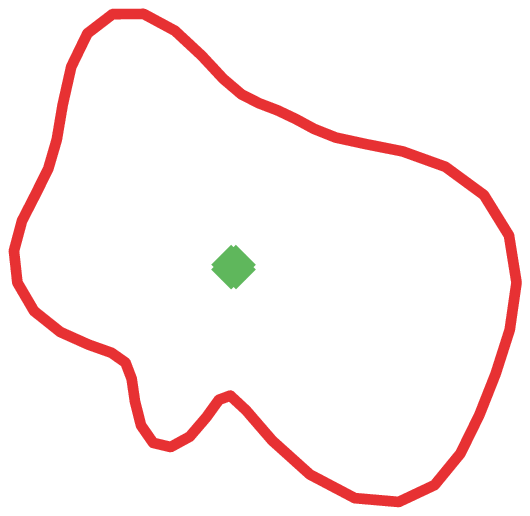}
      \\ 
      \flushleft
      \hspace{0.5cm}Ground Truth: Rectangle \\ 
      \hspace{0.5cm}Bayesian: Plus \\ 
     \hspace{0.5cm}NN: \textbf{Rectangle} \\ 
    \end{minipage}
    & 
    \begin{minipage}{.25\textwidth}
    \flushleft
      \hspace{0.5cm}\includegraphics[trim= 0 0 0 -10,clip,scale=0.5]{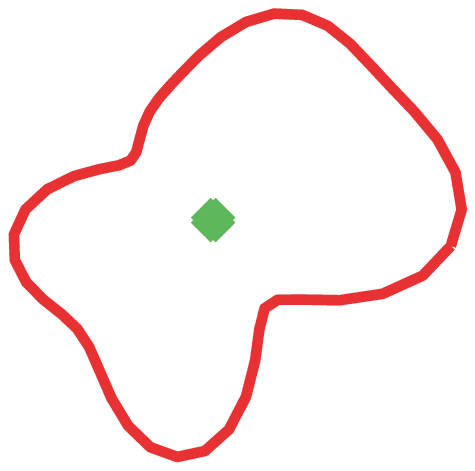}
      \\ 
      \flushleft
      \hspace{0.5cm}Ground Truth: Rectangle \\ 
      \hspace{0.5cm}Bayesian: Triangle \\ 
      \hspace{0.5cm}NN: \textbf{Rectangle} \\ 
    \end{minipage}
    \\ \hline\hline
  \end{tabular}

\end{table*}

A common problem encountered during the training phase of an NN is over-fitting to the training data. 
This leads to degradation in the generalization capabilities of the NN to other data sets. 
With this in mind, we briefly examined the effect of $\mathcal{L}_2$ regularization on the performance of NN-feature and NN-contour. 
For both of the NN-based algorithms, regularization did not improve the accuracy rate of the corresponding model. 

It can be concluded that utilization of an NN in the classifier yields more robust performance especially for irregular contour estimates, which is frequently encountered in real-world scenarios when the tracked object is occluded by its surroundings. 
However, the simplicity of the Bayesian classifier shows its strength at the execution time by being approximately three times faster than the proposed method. 

\section{Conclusion and Future Work}
\label{conclusion}
Classifying objects based on their extent estimates is still at very early ages in the ETT literature. 
In this work, we propose to combine a well-known deep learning algorithm with a GP based extended target tracker to classify the type of dynamic objects. 
To this end, various shape features and different structures of NNs are examined. 
The performance of the suggested algorithm is comparatively demonstrated against a Bayesian classifier. 
NN-based classifier shows superior performance compared to the Bayesian classifier. 
As future work, the algorithm will be tested on real data sets considering several applications such as identification of biological cells and annotating agents in urban driving environment.

\bibliographystyle{IEEEtran}
\bibliography{IEEEabrv,refs}
\end{document}